\documentstyle[preprint,aps,epsf]{revtex}   
\begin{document}

\hfuzz=100pt
\hbadness=10000


\title{\bf Stable unidimensional arrays of coherent strained islands}

\author{Nicolas Combe\footnote{Corresponding author. Fax: +33
    472432648, Email adress : ncombe@dpm.univ-lyon1.fr}, Pablo Jensen and Jean-Louis Barrat}
\address{D\'epartement de Physique des
Mat\'eriaux, UMR CNRS 5586, Universit\'e Claude Bernard Lyon-1, 69622
Villeurbanne Cedex, FRANCE}

\maketitle

\begin{abstract}
  
  We investigate the equilibrium properties of arrays of coherent
  strained islands in heteroepitaxial thin films of bidimensional
  materials.  The model we use takes into account only three essential
  ingredients : surface energies, elastic energies of the film and of
  the substrate and interaction energies between islands via the
  substrate. Using numerical simulations for a simple Lennard-Jones
  solid, we can assess the validity of the analytical expressions used
  to describe each of these contributions. A simple analytical
  expression is obtained for the total energy of the system.
  Minimizing this energy, we show that arrays of coherent islands can
  exist as stable configurations. Even in this simple approach, the
  quantitative results turn out to be very sensitive to some details
  of the surface energy.
\end{abstract}
\vspace{5mm}

\narrowtext

\section{Introduction}
\label{intro}

Manufacturing  arrays of three dimensional nanoscale islands is a
field  of intense research,  with  many potential  optoelectronic
applications ('quantum dots') \cite{levi}. The production of these
structure represents a difficult challenge. One possible way to
proceed is by spontaneous formation \cite{venezuela} of coherent
(i.e.\ dislocation free) strained islands in heteroepitaxy : this
approach represents an elegant and simple way to produce quantum
dots. For instance, when  growing a layer  of InGaAs on GaAs(001)
\cite{guha} or of SiGe on Si(001) \cite{mo_savage,rloo},
spontaneous formation of three dimensional coherent islands is
observed after depositing  a few  layers.  However, our
understanding of the mechanisms that lead to island formation from
strained layers is still incomplete, and even the equilibrium
properties of such structures are not well understood
\cite{tersoff_tromp,shchukin_revue}.

A flat strained layer is {\it unstable}
\cite{grinfeld,srolovitz,spencer} against island formation. The
driving force of this instability is the elastic energy gain due to
the relaxation of the stress at the top of the islands. Basically, the
elastic energy gain for the creation of islands is proportional to the
volume $V$ of the islands (per unit area of an island). The surface
energy loss, on the other hand, scales with the volume as $V^{2/3}$.
Hence it is clear that, above some critical volume, the growth of the
island will be energetically favored. In this simple description, the
equilibrium state of the system consists in a single very big island.

 Experiments, however,  reveal a different behavior.
Regularly spaced islands with a narrow size distribution
\cite{ross} are observed on the substrate, sometimes above a thin
wetting layer. To explain the difference between the simple
argument and the observations, several authors have suggested that
the density of islands is fixed by the kinetics of the growth
process \cite{priester,wang_kratzer}. However, some systems of
islands are not affected by reasonable annealing
\cite{reponse_lee}. Either the driving force becomes very small so
that the system never reaches its equilibrium state, or there is a
stable or a metastable state corresponding to a given island
density. Indeed, Shchukin et al. \cite{shchukin} have argued that
coherent arrays of tridimensional islands may be thermodynamically
stable, when  taking into account elastic energy, facets energy,
surface stress and the contribution of the edges.

In this paper, we study the equilibrium properties of a
bidimensional strained layer. Our aim is to gain an understanding
of the structure by using a minimal model, with the simplest
possible ingredients. To describe the energetics of the layer,
 we only take into account surface energies, interactions between
islands, and the elastic energy of the substrate and the strained
layer.  We have also considered the possibility that the islands
grow on a wetted surface so that our analysis is also relevant to
the Stranski-Krastanov growth mode. Some of the terms that
contribute to the energy cannot be computed analytically, or only
with uncontrolled approximations. When this is the case, we use
simulation of a simple model (Lennard-Jones interactions) to
investigate the approximations done in the analytical calculation.

Minimizing the total energy of an array of islands, we find that,
using only  these simple ingredients,  stable arrays of islands
are predicted in a range of parameter values. The size and density
of the islands at equilibrium and the height of the wetting layer
are also obtained.

In the first  part of the paper, we define the system we study, and
the model used in the simulations.   In the second part, we will
precisely review all the terms that enter the energy calculation.
Elastic terms are calculated using  linear elasticity theory,
supplemented by atomic simulation results. Results are presented
in section~\ref{result}, and briefly discussed in section~\ref{discussion}

\section{Definition of the system and simulations. }

\subsection{The system}
\begin{figure}
\centerline{\epsfxsize=12cm
\epsfbox{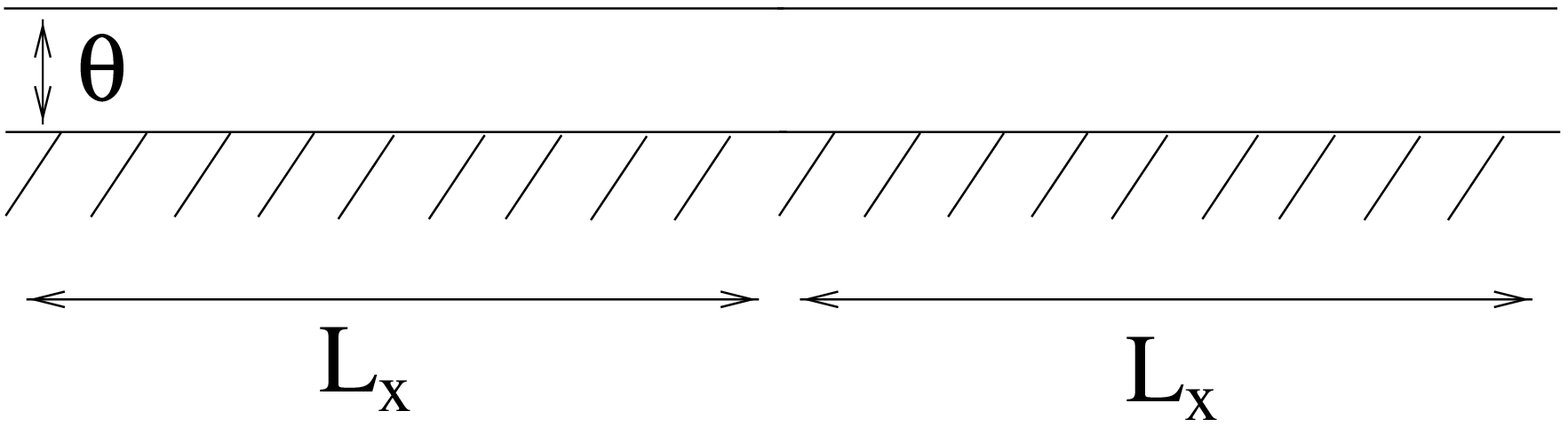}
}
\centerline{a)}
\centerline{\epsfxsize=12cm
\epsfbox{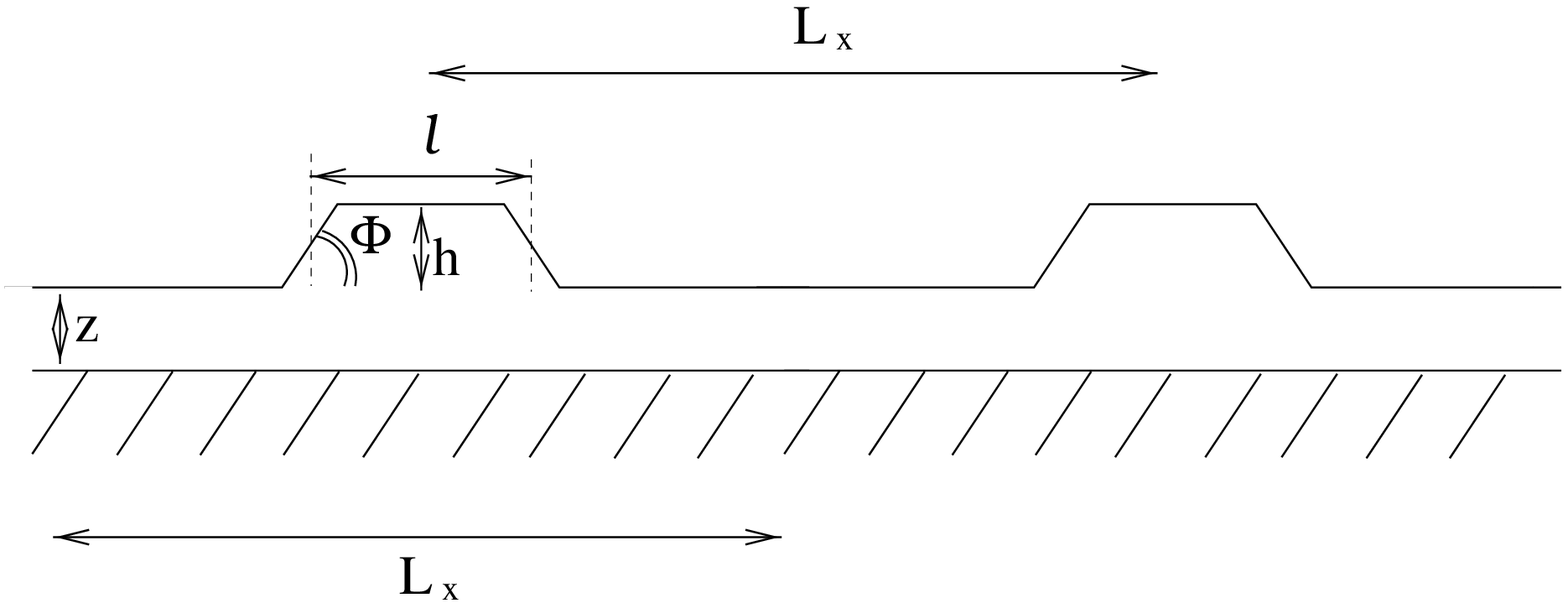}
}
\centerline{b)}
\caption{a) Flat strained layer.\ b) Strained layer
  containing islands.}
\label{system}
\end{figure}

In this paper we compare the energies of a strained layer
containing an array of coherent islands, with that a flat strained
layer of the same volume. For the sake of simplicity, we restrict
ourselves to a bidimensional geometry. In figure~\ref{system},
the two situations for which we want to compare the energies are
described schematically, and the geometric parameters relevant to
describe each configuration are introduced.
In both situations (uniformly strained layer or regular array of
islands),  $\theta$  is the amount of deposited particles per unit
of length. $\theta$ is also the height of the uniform layer in
figure~\ref{system}(a).  In figure~\ref{system}(b), $h$ and
$\ell$ are the height and width of the islands
\cite{tersoff_tromp,tersoff_legoues} so that the volume of an
island is $V=h \ell$. We choose the angle $\Phi$ equal to $\pi/3$.  $z$ is the height of the wetting layer. The
distance between the islands is $L_x$. Note that an infinite value
of $L_x$ corresponds to a substrate with a single island. Since
 the amount of  matter deposited  is the same in both configurations
(a) and (b), we  have
\begin{equation}
\theta L_x = z L_x + h\ell  \label{conserv_mat}
\end{equation}
Our aim is to calculate the difference of energy $\Delta E$
between the two configurations shown in  Fig.~\ref{system}.
$\Delta E$ is a function of $\ell,h,Lx,z$ and $\theta$. $\theta$,
which is the amount of adatoms deposited on the surface, is a
controlled, fixed parameter. Since Eq.~\ref{conserv_mat} connects
the four variables $\ell,h, L_x$ and $z$, $\Delta E$ is a function
of  only three independent variables. We choose $V,r$ and $z$ as
independent variables where, $V=h\ell$ is the volume of an island
and $r=h/\ell$ its aspect ratio.

 By minimizing the difference of
energy per particle $\Delta E / \theta L_x$ at constant $\theta$
and for a given $V$ with respect to $z$ and $r$, we obtain $\Delta E
/\theta L_x$ as a function of the volume $V$ of an island at
given coverage. We can then locate the equilibrium states by
minimizing with respect to $V$.

\subsection{Simulations }
\vspace{.4cm}
\label{simul}

To evaluate some elastic terms in the difference of energy $\Delta
E$ or to check some assumptions in our analytical calculations, we
have performed simulations with a `Lennard Jones material'. In
this material, particles interact via a Lennard-Jones potential.
The Lennard-Jones interaction potential is given by:
\begin{equation}
 V(r) = 4 \epsilon \left[
 \left(\frac{\sigma}{r}\right)^{12} - \left(\frac{\sigma}{r}\right)^{6} \right]
\end{equation}
Again for the sake of simplicity,  we  consider only first
neighbors interactions. This implies that the strain at the
surface is identical to the strain present in the bulk material
(no surface relaxation), so that there is no  excess strain or
stress associated to  the surface. These model materials therefore
have no surface stress.  We use a bidimensional description and
assume that the strain free material (substrate and layer) has a
triangular lattice structure. Since there are two types of
particles, 3 couples of parameters are needed to describe
interactions between atoms in our system : substrate-substrate
($\epsilon_{ss}$, $\sigma_{ss}$), layer-layer ($\epsilon_{ll}$,
$\sigma_{ll}$) and substrate-layer ($\epsilon_{ls}$,
$\sigma_{ls}$). We take the lattice spacing of the
substrate as the unit of length, so that we have only to specify
the misfit $m_0$ between the two materials. We assume that the value of
$\sigma_{ls}$ is the average of $\sigma_{ss}$ and $\sigma_{ll}$.
So that we have\footnote{This choice of $\sigma_{ss}$ gives an
equilibrium
  distance between particles of the substrate equal to $1$.}:
\begin{eqnarray}
\sigma_{ss} &=& \frac{1}{\sqrt[6]{2}}  \\
\sigma_{ll} &=& \sigma_{ss}(1 - m_0)\\
\sigma_{ls} &=& \frac{\sigma_{ss} + \sigma_{ll}}{2}
\end{eqnarray}
In our simulations, we choose the values of the bonding energies
and of the misfit to be: $\epsilon_{ss}=0.058 eV$,
$\epsilon_{sl}=0.06 eV$, $\epsilon_{ll}=0.04 eV$ and $m_0 = 0.06$.

In order to limit the system size,
 we assume that only the atoms of the substrate in the first
$n_{sub}$ layers close to the surface can relax. Since the Green
function \cite{landau_elasticity} of a semi infinite system
decreases as $1/r$ where $r$ is the distance between an applied
force on the surface and a given point in the system, the effect
of limiting the depth of the substrate is basically to screen the
horizontal forces beyond a distance $n_{sub}$. Hence, two islands
at a distance more than $n_{sub}$ do not interact in our
simulations. We use a substrate of typical depth  $60$ (in substrate
lattice spacing units) or
more. We will check in part~\ref{subsect_interaction} that at this
distance, interaction forces between islands become indeed
negligible.

For a given topology (islands or strained layer), our simulation
consists in  relaxing all atomic positions towards the nearest
energy minimum, using a standard conjugate gradient algorithm. We
are then able to evaluate the elastic energy of the system by
subtracting the sum of the equilibrium bonding energy for each
pair of particles from the total energy of the system.

Remembering that interactions are only present between nearest
neighbors, we can compute the elastic constants and surface
energies of the material (at zero temperature).  The surface
energies of the substrate per surface atoms is
$\gamma_{s}=\epsilon_{ss}$, that of the epitaxial layer
$\gamma_{l}=\epsilon_{ll}$.  The interface energy is
$\gamma_{sl}=(\epsilon_{ss}+\epsilon_{ll})-2 \epsilon_{sl}$ and
the Young modulus of a bidimensional Lennard Jones crystal with a
triangular lattice for the substrate $E_s=\frac{144}{\sqrt{3}
\sqrt[3]{2}} \frac{\epsilon_{ss}}{\sigma_{ss}^2}$ and for the
layer $ E_l=\frac{144}{\sqrt{3} \sqrt[3]{2}}
\frac{\epsilon_{ll}}{\sigma_{ll}^2}$.  We can also compute the
Poisson ratio which is the same for both  materials : $\nu =
1/2\sqrt{3}$. With the numerical values we use as parameters, the
strained material totally wets the substrate: $\gamma_{s} -
(\gamma_{l}+\gamma_{sl} ) > 0$.

\section{Ingredients}
\label{ingredients}

In this section, we list the
 different contributions to the
difference of energy $\Delta E$ between the two configurations
mentioned in Fig.~\ref{system}. Three essential ingredients are
taken into account: elastic energies, interactions between islands
and surface energies. Each  of these contributions is discussed
below.

\subsection{Relaxation energy}
\label{relax_energ}

The elastic energy per unit length of a film of height $\theta$
with a uniform strain $m_0$ is given by the expression :
\begin{equation}
E_{elastic}^{flat}/L_x = \frac{E_l}{2} m_0^2 \theta
\end{equation}
Where $E_l$ is the Young modulus of the strained material.

The exact expression of the elastic energy of a (substrate+island) system
is, to our knowledge, unknown. Two studies
\cite{Muller_Kern2,Muller_Kern3,ratsch_zangwill} propose an
approximate expression for this elastic energy, in the absence of a
wetting layer ($z=0$). In both studies, the elastic energy of an
island is written as :
\begin{equation}
E_{elastic}^{island} = \frac{E_l}{2} m_0^2 R(r) V. \label{eq_relax}
\end{equation}
Here $V$ is the volume of the island and $R(r)$ is a dimensionless
factor determined by the shape of the island and  the elastic
constants  of both the substrate and layer materials. These
studies show that the  shape factor depends essentially of the
aspect ratio $r$ of the island, and only weakly of the ratios
between elastic constants.
 This shape factor takes into
account both island and substrate contributions to the elastic
energy.

\begin{figure}
\centerline{\epsfxsize=12cm
\epsfbox{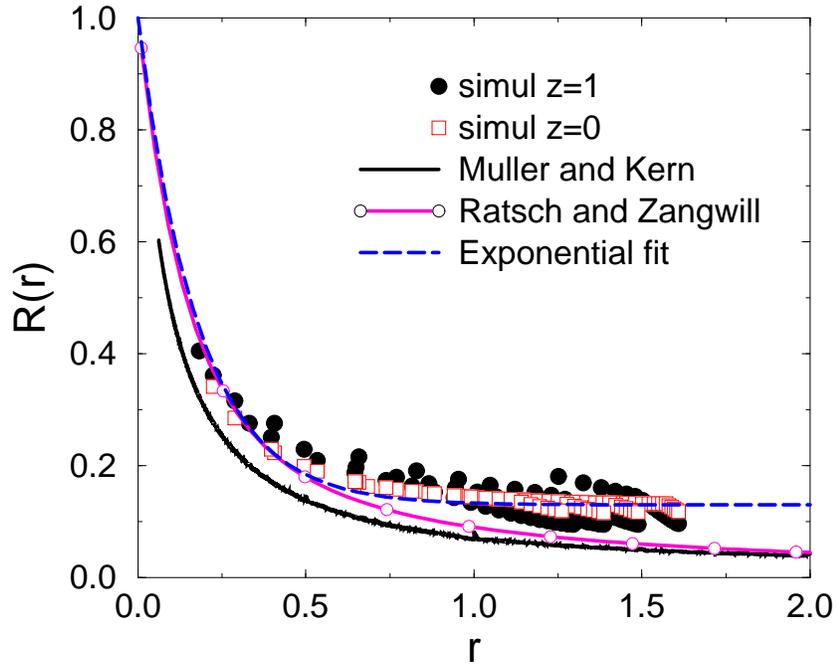}
}
\caption{Shape factor $R(r)$ as a function of the aspect ratio $r$. The solid line has been obtained from the equation : $R(r)=0.13
  +0.87*\exp(-r/0.18)$.  We have also compared our simulations with
  the predictions of Kern and M\"uller and Ratsch and Zangwill.}
\label{relaxation}
\end{figure}

Fig.~\ref{relaxation} presents the shape factor $R(r)$ calculated from
our simulations for the peculiar case of islands with a $\pi/3$ contact
angle. $R(r)$ goes to 1 when $r\rightarrow 0$ : a very flat
island can not relax its elastic energy. As $r$ increases, $R(r)$
decreases, since atoms in the islands can relax as soon they are far
enough from the substrate. In our simulation, we cannot evaluate the
shape factor $R(r)$ for aspect ratio greater than
$r_{max}=\tan(\phi)\approx 1.7$ due to the choice of the islands
shape.

For the numerical calculations of Sect.~\ref{result}, we need a
accurate analytical expression of $R(r)$.  $R(r)$ physically
corresponds to the elastic energy relaxation due to the creation of an
island, it will then appear in the only negative term of the
expression of $\Delta E$ : our final result will thus be sensitive to this
function.  To obtain an analytical expression for $R(r)$, we have
fitted the results by an exponential : we find $R(r)=0.13
+0.87*\exp(-r/0.18)$.  We have checked that neither the expression of
Ratsch and Zangwill \cite{ratsch_zangwill} (which contains a free
parameter) nor the expression of Kern and M\"uller
\cite{Muller_Kern2,Muller_Kern3} (no free parameter) allow to fit our
simulation results (Fig.~\ref{relaxation}). However, this disagreement
may be due to the difference of contact angle between their islands
and ours \cite{wong_thouless} : Ratsch and Zangwill and Kern and Muller
islands have a contact angle $\pi/2$ with the substrate,
whereas our islands have a $\pi/3$ one.  Moreover, our simulation also show
that the shape factor does not significantly change with the height of
the wetting layer.

\subsection{Interaction between Islands}
\label{subsect_interaction}

\begin{figure}
\centerline{\epsfxsize=12cm
\epsfbox{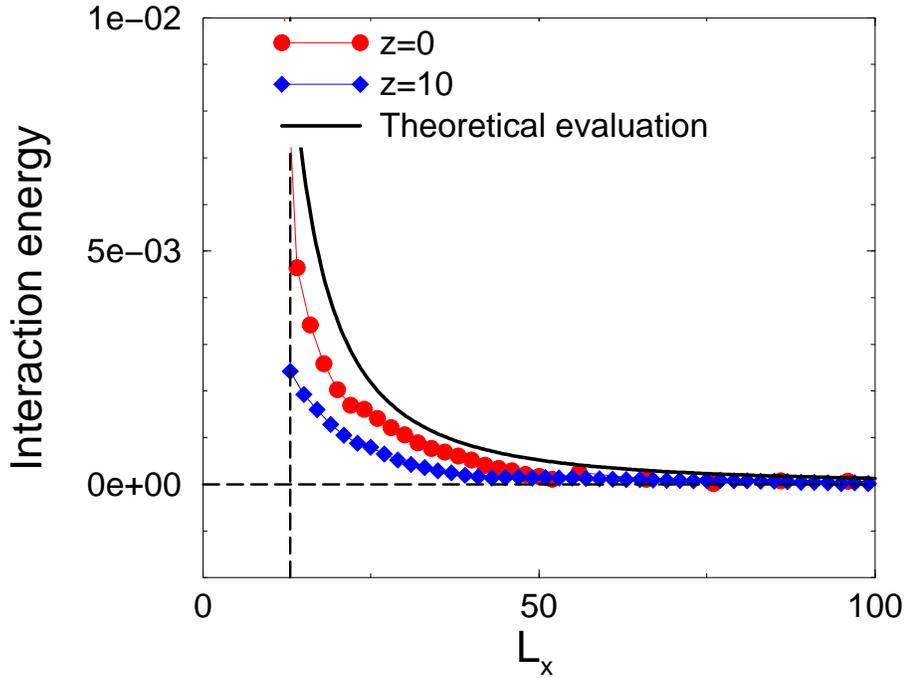}
}
\caption{Interaction Energy between two islands as a function of the
  distance between islands. We have drawn two curves obtained by
  simulations for different heights of the wetting layer ($z=0$ and
  $z=10$), and the curve obtained from Eq.~\ref{e_interactions}
  without any adjustable parameter. The vertical
  dashed line corresponds to the two islands touching.
  Values of $n_{sub}$ used for the calculation of the curve are
  $n_{sub}=240$ for $z=0$ and $n_{sub}=90$ for $z=10$.}
\label{interaction}
\end{figure}

To evaluate the interaction energies between islands, we use the same
formalism, the same approximations and the same islands shapes (cf.
Fig.~\ref{system}) as Tersoff \cite{tersoff_tromp,tersoff_legoues}. We
assume that an island with a shape given by $h(x)$, exerts on the
substrate surface an horizontal force density given by $f_x= \sigma_0
\partial_{x} h(x) $ where $\sigma_0$ is the bulk stress in a layer
uniformly strained to fit the substrate.  We calculate the
displacement $u_x$ of an atom of the substrate surface caused by the
presence of an island at a distance $L_x$ using the Green function
\cite{landau_elasticity} of a semi infinite plane. We then evaluate
the strain $u_{xx}$ caused by the
presence of the island : 
\begin{eqnarray}
u_x&=&\frac{E_l}{\pi E_{s}} m_0 \tan(\phi) \left[
 L_x \ln{ \frac{L_x^2 - (\ell/2+a/2)^2}{L_x^2 - (\ell/2-a/2)^2}} \right. \label{ux_interact} \\
&&+ l \ln{ \frac{(L_x+a/2)^2 - (\ell/2)^2}{(L_x-a/2)^2 - (\ell/2)^2}} \nonumber \\
&& \left. + a \ln{ \frac{(L_x+\ell/2)^2 - (a/2)^2}{(L_x-\ell/2)^2
- (a/2)^2}} \right]
\nonumber \\
u_{xx}&=&\frac{\partial u_x}{\partial L_x} \label{uxx_interact}
\end{eqnarray}
where $a$ is defined by $a=h/\tan(\phi)$ and  we have used
$\sigma_0 = E_l m_0$. $\ell$, $h$ and $\phi$ are defined on
Fig.~\ref{system}. $E_s$ and $E_l$ are respectively
the Young modulus of the substrate and of the strained material.\\
An island located at a distance $L_x$ from another sees a misfit
 $m=m_0-u_{xx}$ different from the original misfit $m_0$.
  From the elastic energy Eq.~\ref{eq_relax} of a single island,
we deduce the interaction energy between two identical islands
separated by  a distance $L_x$:
\begin{equation}
E_{interact}=  E_l R(r) m_0^2 \left[(1-\frac{u_{xx}}{m_0})^2 -1 \right] V
\label{e_interactions}
\end{equation}
where $u_{xx}$ is given by Eqs.~\ref{ux_interact}
and~\ref{uxx_interact}.  A first order asymptotic expansion of Eq.~\ref{ux_interact}
shows that $u_{xx}$ scales as $-1/L_x^2$ for large
$L_x$. Eq.~\ref{e_interactions} then  corresponds  to
 a repulsive interaction between islands. Figure~\ref{interaction}
shows a comparison between the interaction energy of two islands
obtained from simulations using the  Lennard Jones model and
 the result of equation~\ref{e_interactions}. Remembering that there is no
adjustable parameter in Eq.~\ref{e_interactions}, this expression leads to a reasonable estimate  for the
interaction energy. We have neglected the influence of the height
of the wetting layer. A wetting layer, however can be taken into
account in the simulation, and  fig.~\ref{interaction} shows that
 it does not modify the order of magnitude of 
interaction energies. 

Moreover, the main effect of this interaction energy in the whole
expression of $\Delta E$ consists in a strong short range repulsion.
Thus, with or without the presence of a wetting layer, the approximate
expression~\ref{e_interactions} should allow to get the physical
effects of the repulsive interactions between islands.

\subsection{Surface Energy}

We have to calculate the surface energy difference between the two
configurations of Fig.~\ref{system}. For a material without surface
stress, and with the elastic terms calculated Sect.~\ref{relax_energ}
already taking into account the strain of the surface, we only have to
consider the strain free surface (i.e.\ atoms are in their equilibrium
position). Following M\"uller and Kern \cite{mull_kern} and
simulations results of Wang {\it et al.}  \cite{wang_kratzer}, the
surface energy of the layer and the interface energy depend on the
height of the layer, and we define $S(z)$ :
\begin{equation}
S(z)=\gamma_{l}(z)+\gamma_{sl}(z)
\end{equation}
$S(z)$ is the surface energy cost to create a wetting layer of height
$z$ on the substrate.  When $z$ tends to infinity, $S(z \to \infty)$
tends to $\gamma_{l}+\gamma_{sl}$ where $\gamma_{l}$ and $\gamma_{sl}$
has been given in Sect.~\ref{simul}. When $z$ tends to $0$, the
wetting layer disappears, and only the substrate is left, so that $S(z
\to 0)$ tends to $\gamma_s$. As suggested by M\"uller and Kern
\cite{mull_kern}, we choose an exponential function for $S(z)$ (valid
for semiconductors and metals). This choice should qualitatively give
the correct behavior for the function $S(z)$.
\begin{equation}
S(z)=\left(\gamma_{s}-\gamma_{l}-\gamma_{sl}\right) e^{-z/z_0} +
  \gamma_{l}+\gamma_{sl}
\label{S(z)}
\end{equation}
And following M\"uller and Thomas \cite{muller_thomas}, we define
surface energies as :
\begin{eqnarray}
\gamma_{l}(z)&=&\gamma_{l} \left(1- e^{-z/z_0}\right)  \\
\gamma_{sl}(z)&=&\left(\gamma_{sl}-\gamma_{s}\right) \left(1- e^{-z/z_0}\right) +
  \gamma_{s}
\end{eqnarray}
Where $z_0$ is a parameter comprised basically between $1$ and
$5$. We have chosen $z_0 = 3$ in this work. This choice of the
screening distance $z_0$,  and the choice made for  the function
$S(z)$ will be discussed in Sect.~\ref{discussion}. We are now
able to write the difference of surface energy between the two
systems of Fig.~\ref{system}.
\begin{eqnarray}
\Delta E_{surface} &=& \left[ S(z) - S(\theta) \right] [L_x - \ell - a] \nonumber\\
                   & & + \gamma_{l}(z+h) \left[ \ell-a + 2
                     \frac{h}{sin(\phi)} \right]\nonumber \\
                   & & + \gamma_{sl}(z+h)\left[\ell+a\right]
                   \nonumber \\
                   & & - S(\theta) [\ell+a] \label{surface}
\end{eqnarray}
With $a=h/\tan(\phi)$.\\
The first term on the right side of Eq.~\ref{surface} is the
surface energy difference for a film of length $L_x-\ell-a$ when
changing the thickness from  $\theta$ to  $z$. The last three
terms correspond to the surface energy difference between a flat
film (width $\ell+a$) and the island (height $h$ and width $\ell$). We
take $\phi=\pi/3$ so that for a triangular lattice, island sides
surfaces are equivalent to top surfaces.

\section{Results}
\label{result}

Using Eq.~\ref{e_interactions} and~\ref{surface}, we have all the
ingredients to write the expression of $\Delta E$.
\begin{eqnarray}
\Delta E &=& \frac{E}{2}  m_0^2 \left[ R(r)(1-\frac{u_{xx}}{m_0})^2 -1 \right]
V \nonumber\\
         & & + \left[ S(z) - S(\theta) \right] [Lx - \ell - a] \nonumber\\
         & & + \gamma_{l}(z+h) \left[ \ell-a + 2 \frac{h}{sin(\phi)}
         \right] \nonumber \\
         & & + \gamma_{sl}(z+h)[\ell+a] \nonumber \\
         & & - S(\theta) [\ell+a]\label{dE}
\end{eqnarray}
Where $u_{xx}$ is given by Eqs.~\ref{ux_interact} and~\ref{uxx_interact}.
Using variables $V,r$ and $z$, and writing the energy per particle, we
obtain :
\begin{eqnarray}
\frac{\Delta E}{\theta L_x} &=& \frac{E}{2}  m_0^2 \left[ R(r)(1-\frac{u_{xx}}{m_0})^2 -1
\right] \frac{\theta-z}{\theta} \nonumber  \\
& & + \frac{S(z) - S(\theta)}{\theta} \left[ 1- \frac{
    \sqrt{V/r} + \sqrt{rV} /\tan(\phi)}{V} (\theta -z) \right]  \nonumber \\
& & + \gamma_{l}(z+\sqrt{rV}) \frac{\theta-z}{V \theta} \left[
  \sqrt{V/r}-\frac{\sqrt{rV}}{\tan(\phi)} + 2 \frac{\sqrt{rV}}{sin(\phi)}
\right] \nonumber \\
& & +
\gamma_{sl}(z+\sqrt{rV})\left[\sqrt{V/r}+\frac{\sqrt{rV}}{\tan(\phi)}\right]\frac{\theta
  -z}{\theta V} \nonumber \\
& & - S(\theta) \frac{\theta-z}{V \theta}
\left[\sqrt{V/r}+\frac{\sqrt{rV}}{\tan(\phi)} \right]
\label{Etotal}
\end{eqnarray}
Where we have used $h=\sqrt{rV}$, $\ell=\sqrt{V/r}$ and
$L_x=\frac{V}{\theta-z}$ from Eq.~\ref{conserv_mat}. 
Each term of equation~\ref{Etotal} has been derived from the
simulations. But we now assume that this equation is valid for any
value of the parameters and is independent of the details of the
simulations (for example, the value of $n_{sub}$ is not meaningful in
equation~\ref{Etotal}).  \\
For $\theta$ and
$V$ given, we minimize Eq.~\ref{Etotal} compared to the variables $r$
and $z$ using conditions $0 \leq z \leq
\theta$ and $ (\ell+a) \leq L_x$ (islands cannot overlap).\\
Figure~\ref{mini} presents the results of this minimization procedure:
each curve corresponds to a fixed value of $\theta$.  For large enough
coverages, a minimum in the curve is obtained at small volumes,
corresponding to the stability of an array of islands. For greater
volumes, the function $\Delta E/ \theta L_x$ increases up to a local
maximum, and then decreases up to a {\it horizontal asymptote} for
infinite volume. This decrease is not shown on Fig.~\ref{mini}
because of the choice of the axis scale.  In the case of
Fig.~\ref{mini}, for each curve presented, the horizontal asymptote
has a greater energy level than the local minimum at small volume, so
that these minima correspond to a {\it stable} configuration.

\begin{figure}
\centerline{\epsfxsize=12cm
\epsfbox{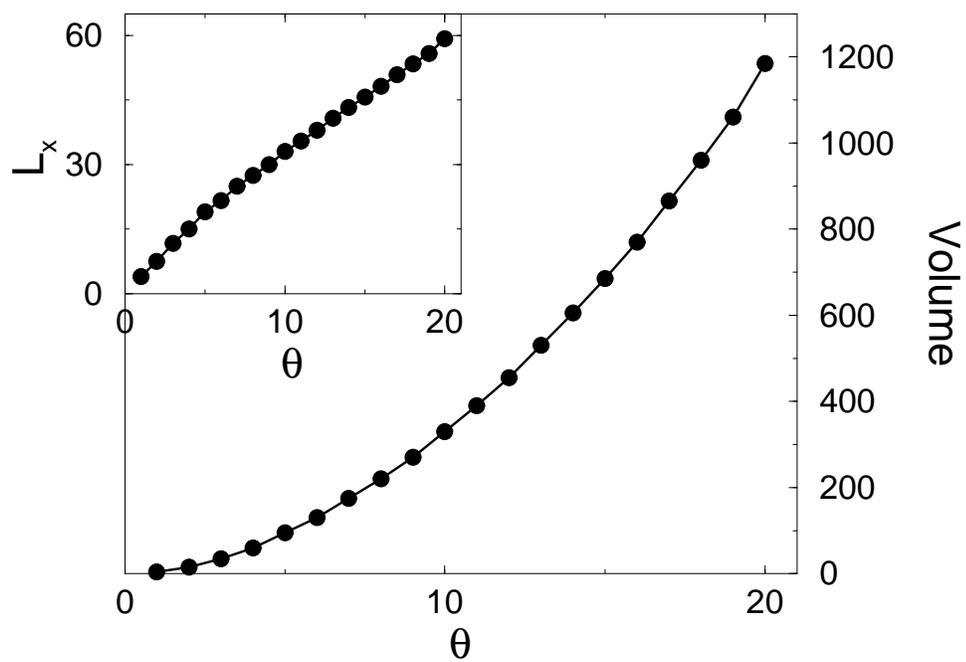}
}
\caption{Variation of the volume of the islands and of the distance
  between islands at the minimum of energy as a function of the coverage
 $\theta$.}
\label{volume}
\end{figure}

\begin{figure}
\centerline{\epsfxsize=12cm
\epsfbox{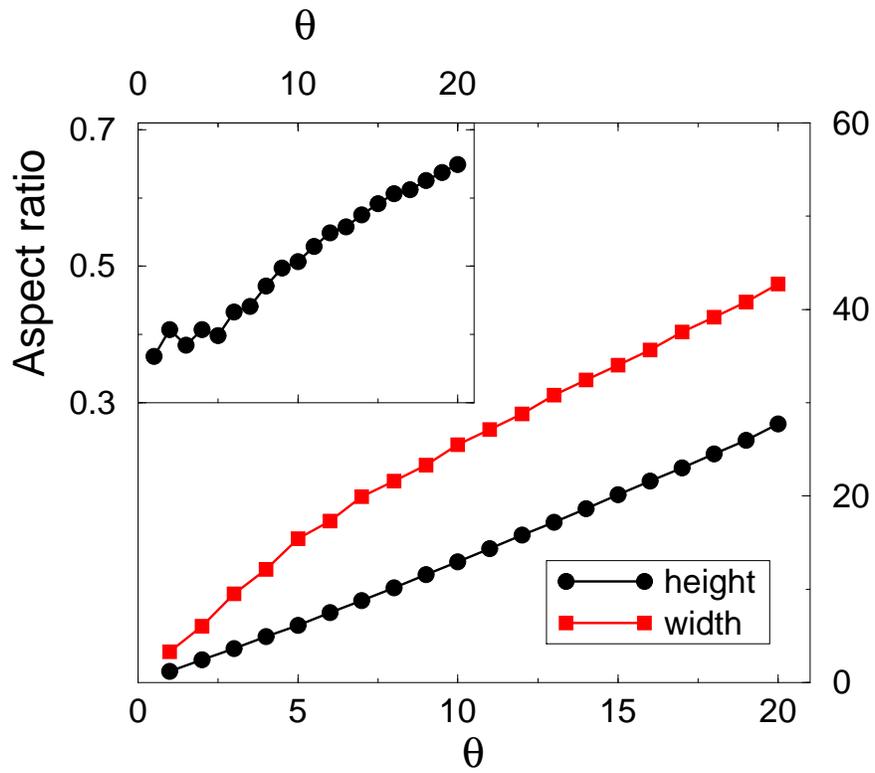}
}
\caption{Variation of the island  height and width at the
minimum of energy as a function of the coverage
 $\theta$. In the inset, we plot the corresponding island aspect ratio as a function
 of $\theta$.}
\label{hauteur}
\end{figure}

With the numerical values we use for the parameters, the flat layer is
unstable for a coverage $\theta$ larger than $1$. The critical
thickness is thus lower than $1$. However, for $\theta=1$, the volume
of the stable islands is only $4$ and the height and width is about
$1.2$ and $3.4$. Thus, if these values were valid in an experience, the
measured critical thickness would be greater than $1$ since these
islands would be of about the same size as the thermal perturbations of
the flat surface.
 For a coverage
greater than $1$, the energy is minimum for an island volume ranging
from $4$ to $1200$ particles, depending on the coverage.  
Fig.~\ref{volume} shows the evolution of the volume at the minimum of
energy $\Delta E/ \theta L_x$ as a function of the coverage $\theta$.
This volume is an increasing function of the coverage.  At the
minimum, we find that $z=0$, i.e.\ there is no wetting layer. On
Fig.~\ref{volume}, we also show the distance $L_x$ between islands in
the minimum energy configuration. The island density decreases with
the coverage $\theta$.  Fig.~\ref{hauteur} shows
the island height, width and aspect ratio as a function of the
coverage at the minimum of energy.  The aspect ratio remains small
(lower than $0.7$) for all the coverages.

\section{Discussion}
\label{discussion}

Our study  shows that the existence of stable arrays of islands
can be explained using relatively simple ingredients. The
equilibrium state depends on the value of the coverage, in
qualitative agreement with experimental observations \cite{rloo}.
Typical  volumes for the islands are of the order of  500
particles, and typical widths are in the range from 20 to 40 (in
substrate lattice spacing units). This corresponds to islands of
width from 4.7 nm to 9.4 nm,  using the silicon lattice spacing. These widths
are in good agreement with measured widths of ``huts'' in the Ge on
Si (001) experiments \cite{mo_savage}. 

Moreover, we find that there is no wetting layer at equilibrium.
System such as Pd on Cu \cite{abel} or InGaAs on InP \cite{solere}
exhibit this type of behavior and even atoms of the substrate go in
the islands.

\begin{figure}
\centerline{\epsfxsize=12cm
\epsfbox{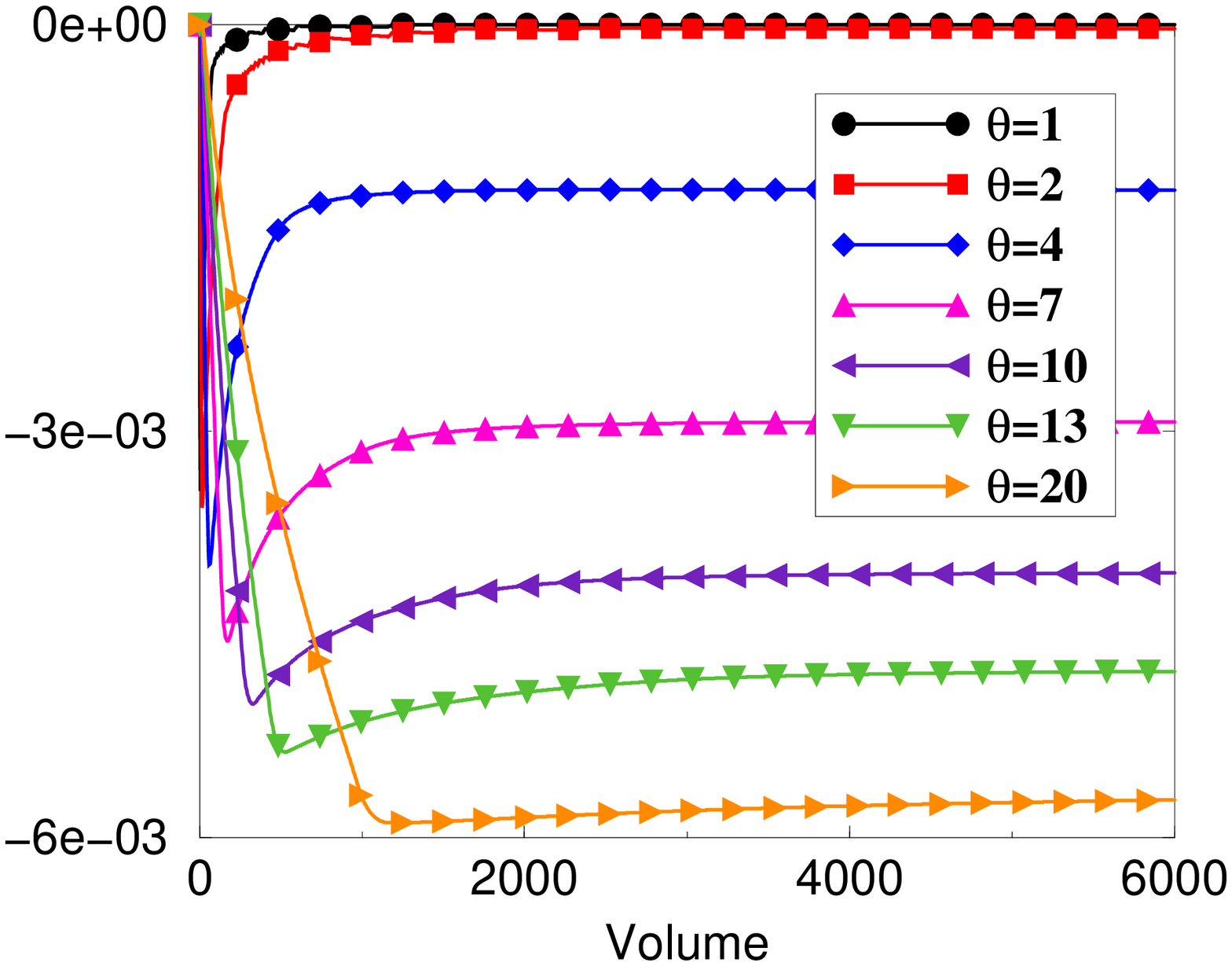}
}
\caption{Variation of Energy per particle as a function of the volume
  for a given value of the coverage $\theta$. For each curve, only
 one point out of 50 has been precised in an aim to keep the figure
 clear.
}
\label{mini}
\end{figure}

\begin{figure}
\centerline{\epsfxsize=12cm
\epsfbox{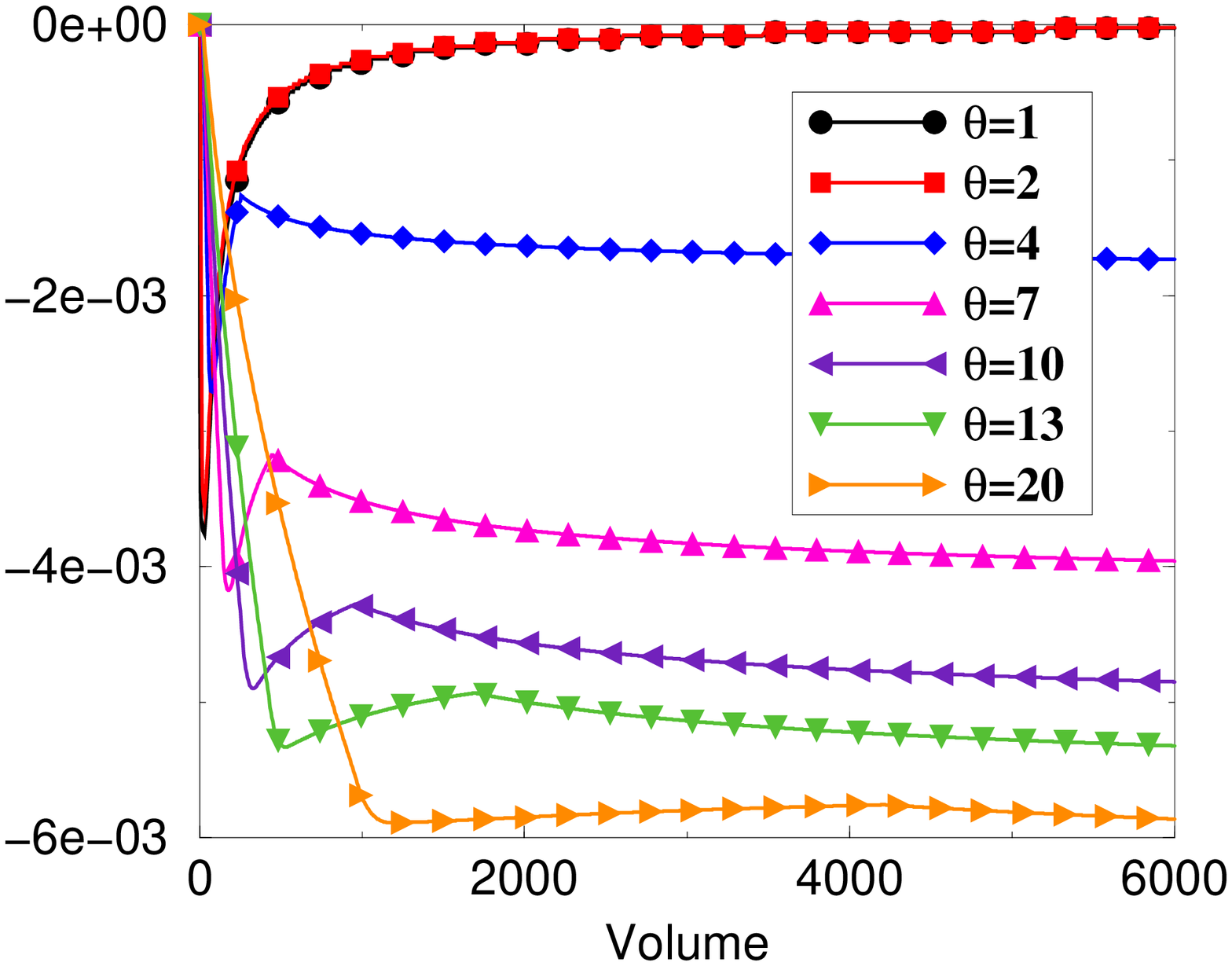}
}
\caption{Variation of Energy per particle as a function of the volume
  for a given value of the coverage $\theta$ with $S(z)$ given by
  Eq.~\ref{S21} and~\ref{S22}. For each curve, only
 one point out of 50 has been precised in an aim to keep the figure
 clear.}
\label{mini2}
\end{figure}

To investigate the sensitivity of our results to the choice of the
function 
 $S(z)$ (equation~\ref{S(z)}), we use the following definition of $S(z)$ :
\begin{eqnarray}
S(z) &= \gamma_{s}- \frac{z(\gamma_{s}-\gamma_{l}-\gamma_{sl})}{3.}  &
\mbox{   for }  z \leq 3 \label{S21} \\
S(z) &= \gamma_{l}+\gamma_{sl} & \mbox{   for } z \geq 3 \label{S22}
\end{eqnarray}
Fig.~\ref{mini2} presents the result of the minimization of
Eq.~\ref{Etotal} with this new expression of $S(z)$. 
Fig.~\ref{mini2} still shows minima but these are metastable and the energy barrier
to leave them is smaller than in Fig.~\ref{mini}.
We also find that our results are sensitive to the value of $z_0$. 

Thus, comparing results of Fig.~\ref{mini} and~\ref{mini2} could
suggest that our model
is very sensitive to the choice of the function $S(z)$.
Nevertheless, this sensitivity is mainly quantitative :  the function $\Delta E/\theta L_x$
also shows a decrease at very large volume (not shown on Fig.~\ref{mini}).
Therefore, changing the function $S(z)$ does not change the
qualitative behavior of the function $\Delta E/ \theta L_x$ : curves
always present a local minimum at small volumes, a local maximum at
intermediate values, and then a decrease to an horizontal asymptote.

Concerning the quantitative disagreement between Fig.~\ref{mini}
and~\ref{mini2}, our model requires the minimization of a function of
several variables. The presence of smooth local minima of this
function may introduce a relative sensitivity to the input quantities.
A plot of the functions described by equations~\ref{S(z)}
and~\ref{S22} shows that the difference between these is not 
negligible. Finally, taking the derivative of
expression~\ref{Etotal} with respect to $V$, we note that the value of
$V$ minimizing Eq.~\ref{Etotal} depends on the value of $dS(z)/dz$.
Unlike equation~\ref{S(z)}, equation~\ref{S22} has a zero derivative
for $z$ greater than $3$ : this unphysical behavior may greatly
affect the results of the minimization of $\Delta E/ \theta L_x$.
  
These results show that a very precise description of
the surface energies would be needed in order to give the
theoretical calculations a predictive character. This was not the aim
of the present paper, but such a 
description may in principle, be obtained from ab-initio calculations
\cite{wang_kratzer}.

Concerning the behavior of $\Delta E/\theta L_x$ at large volumes,
curves of Fig.~\ref{mini2} display an energy minimum for an infinite
island volume for large enough coverages ($\theta > 4$). This minimum
is the one predicted by the very simple argument of Sect.~\ref{intro}
: the corresponding state of the system consists in a single very big
island.  At this minimum, the height of the wetting layer is not 0 but
takes the value $3$ for most coverages $\theta$.  As announced, the
curves of Fig.~\ref{mini} also show these minima at very large volumes, but the wetting layer has a height in the range from 1.3
to 1.7 depending on the coverage. Since the thermodynamical force that
drives the system to this equilibrium configuration decreases to zero
when the volume of islands $V$ tends to infinity, it would not be
surprising that the system does not evolve even after reasonable
annealing because of kinetic limitations. Hence, our analysis suggests
that some experimental observations of large islands are the result of
kinetic factors and that some observed islands do not actually
correspond to an energy minimum.

Fig.~\ref{mini} and~\ref{mini2} do not show any nucleation barrier for
islands formation. This is actually a direct consequence of the
minimization of $\Delta E/\theta L_x$ with respect to $r$ and $z$ : we
are looking at the equilibrium states of a strained layer and a
priori, these are different from the states occurring along the growth of the
system. Basically, in our treatment, if the production of islands is
energetically unfavorable, the equilibrium state of the system is the
flat strained layer and the minimum of $\Delta E/\theta L_x$ is
obtained for $z=\theta$ ($L_x$ has then an infinite value so
that equation~\ref{conserv_mat} remains valid).  Thus, by definition,
reported values of Fig.~\ref{mini} and~\ref{mini2} can not have
positive values and no nucleation barrier can appear in these figures.
Nevertheless, during the growth of a real strained layer, we expect
that a nucleation barrier drives the islands formation. Indeed, before
the formation of large islands, the system has to create smaller one
which would not correspond to equilibrium states described by
Fig.~\ref{mini} and~\ref{mini2}. The nucleation barrier depends on the
kinetic pathway followed by the system to reach its equilibrium state
: thus, the determination of the energy barrier demands a kinetic
study of the transition.

To check the robustness of our model, we performed few tests changing
the values $\epsilon_{12}$ and $\epsilon_{22}$ of the bonding
energies. We have changed the values of the ratio
$\epsilon_{12}/\epsilon_{11}$ from 0.9 to 1.21 (this ratio was fixed
at 1.03 previously) and the ratio $\epsilon_{22}/\epsilon_{11}$ from
0.52 to 0.86 (0.66 previously) : we find that the qualitative behavior
is not affected. Curves always present minima at small volumes with no
wetting layer, a local maxima and then a decrease up to an horizontal
asymptote.

The observations of Ref.~\cite{ross_tersoff} shows that ``huts''
of Germanium on Silicon(001) disappear during the growth and
simultaneously ``domes'' appear. Such result could be interpreted
by an energy landscape such as  the one of Fig.~\ref{mini2} :
`huts' would correspond to a metastable state whereas, `domes'
would correspond to a non equilibrium state. To definitely  answer
this question, one would have to describe very carefully the
system Ge on Si(001).

In Eq.~\ref{Etotal}, we made the assumption that variables
$z,h,\ell$ and $L_x$ are continuous. This is obviously an
approximation. The only alternative, however, would be a fully
atomistic simulation.

 Our analysis does not take into account surface stress in the
elasticity calculation in Sect.~\ref{ingredients}. It is
interesting that nontrivial stable configurations are nevertheless
obtained, showing that  surface stress is not an essential
ingredient in island formation. Surface stress could be included
automatically in the numerical simulations by taking into account
long range interactions between atoms. Another theoretical
challenge would be to take into account the possible formation of
alloys during the growth : this again seems rather difficult at
the analytical level, since it involves taking into account new
elastic coefficients, and new variables such as the volume of the
alloy.

Our work shows that, by combining an analytical approach with some
ingredients extracted from numerical simulations,  a quantitative
study of the stable configurations for a strained layer is
possible. Although we have limited ourselves to a simple two
dimensional geometry and a model Lennard-Jones potential, it seems
possible to extend the approach to more realistic interactions
(e.g. Tersoff' potentials \cite{pot_tersoff} for Si and Ge) and a
three dimensional geometry.\\

\noindent {\bf Acknowledgments}

We are grateful to M. Abel, R. Loo, L. Porte, and Y. Robach for
useful discussions. This work was supported by the P\^ole Scientifique
de Mod\'elisation Num\'erique at ENS-Lyon and by the r\'egion
Rhone-Alpes under the program "contraintes
et r\'eactivit\'e".

\figure

\newpage

\newpage

\newpage
\newpage

\newpage
\newpage
\end{document}